\documentclass[12pt,notitlepage]{revtex4-1}
\usepackage{graphicx}
\usepackage{amsmath}
\usepackage{amsfonts}
\usepackage{amssymb}
\usepackage{epstopdf}

\begin{document}

\title{The nematic phase of a system of long hard rods\\
\small{Invited talk given at the International Congress on Mathematical Physics, Aalborg, August 2012}}

\author{Alessandro Giuliani}

\address{Dip.to di Matematica, Universit\`a di Roma Tre\\
L.go S. L. Murialdo 1, 00146 Roma - Italy}

\begin{abstract}
In this talk I consider a two-dimensional lattice model for liquid crystals consisting of long rods interacting via purely hard core interactions, with two allowed orientations defined by the underlying lattice. I report a rigorous proof of the existence of a nematic phase: by this I mean that at intermediate densities the system exhibits orientational order, either horizontal or vertical, but no positional order. The proof is based on a two-scales cluster expansion: first the system is coarse-grained on a scale comparable with the rods' length; then the resulting effective theory is re-expressed as a contours' model, which can be treated by Pirogov-Sinai methods. 
The talk is based on joint work with Margherita Disertori.
\end{abstract}


\maketitle

\section{Introduction}

A nematic liquid crystal is a phase displayed by certain systems of anisotropic molecules, in which the 
distribution of orientations of the particles is anisotropic, while the distribution of the particles in space is 
homogeneous. A picture to keep in mind is that of a system of cigar-shaped molecules in the 
three-dimensional continuum. The location of each particle in space is characterized by the position of the center 
of mass and by the direction of the main axis of symmetry of the molecule. In a nematic phase, all molecules' 
axes are directed 
approximately in the same way, while the centers are distributed in a random, ``liquid-like" way. More formally,
assuming the microscopic interaction among molecules to be translation and rotation invariant, in a nematic 
phase the underlying rotation symmetry of the system is completely broken, while the translation symmetry remains
unbroken. 

Nematic liquids are physically characterized by peculiar refraction properties, which can be conveniently used to 
filter polarized light, an effect that is extensively used in everyday's life in the liquid crystal displays of our computers and televisions. 
The technological importance of nematic liquids, which became clear already in the first half of the twentieth 
century, has been the boost for a large number of theoretical studies about the microscopic mechanism underlying 
their occurrence. Depending on the specific nature of the system under consideration, the intermolecular forces 
can be either of repulsive or attractive type. In certain suspensions of colloidal particles (e.g., 
aqueous suspensions of the tobacco
mosaic virus) the dominant intermolecular interaction is a screened repulsive electrostatic force.

For such systems, the theory of the microscopic mechanism underlying the spontaneous formation of a nematic 
phase is due to Lars Onsager \cite{O}: roughly speaking, his idea was to 
picture each molecule as a long, thin rod, each characterized by a translational and a rotational 
degree of freedom. Since the intermolecular interaction is repulsive and short ranged, it can be essentially thought 
of as a hard-core repulsion: therefore, the free energy is 
essentially equal to the entropy (i.e. the logarithm of the phase space volume corresponding to the allowed configurations of mutually
non-overlapping molecules), which is the sum of a translational and a rotational 
entropy. At very low densities, so low that the average intermolecular distance is much larger than 
the rod's length $\ell$, every molecule has enough space to rotate freely around its center; therefore, the system is 
an isotropic gas phase. At higher densities, each molecule is surrounded by other molecules at an average  
distance smaller than $\ell$ and, therefore, it is not free to rotate in all directions. In some cases, it may be 
favorable for the molecules to align spontaneously: it may be convenient for the system to substantially 
reduce its orientational entropy, the loss being compensated by a much larger gain of translational entropy.
To quantify this effect, Onsager computed an approximate equation of state by truncating at second order the 
virial expansion and by computing explicitly the first two Mayer's coefficients. 
Then he expressed the equation of state as a functional of the distribution function of the rods' orientations and verified that for certain geometric shapes of the molecules there exists a non vanishing interval of densities
where the free energy of the system is lowest for an inhomogeneous distribution of the rods' orientations. 

Onsager's theory is the first example of an {\it entropy-driven} phase transition, i.e. an ordering
transition induced by the competition of two entropic effects, each of which would separately like to make the 
system as disordered as possible. The mechanism he proposed is often refereed to as the {\it Onsager's excluded 
volume effect}. Since then, there have been several attempts to prove in concrete models the correctness of this 
proposal, whose rigorous understanding is still an open problem  in the context of three-dimensional systems with continuous translational and 
rotational symmetries. On the other hand, there are indications of the validity of the Onsager's mechanism 
in the simpler context of long hard rods systems on $\mathbb{Z}^d$ or on $\mathbb{R}^d$
with {\it discrete} orientations. These can be thought of as toy models for lattice liquid crystals, in which the 
continuous rotational symmetry is replaced by a discrete one, much easier to treat; in this simplified context, we shall 
refer to a nematic-like phase (or nematic tout court) as a phase that breaks the discrete rotational symmetry, 
but not the translational one. In the following we shall briefly review, from a mathematical physics perspective,  
some important previous results concerning such systems. We shall restrict to systems where 
a hard core repulsion among rods is present and the set of allowed orientations is discrete; 
the case where the rod-rod interaction is of attractive nature and continuous rotational symmetry is retained 
is another (very interesting) story, which we will not review here, see \cite{AZ,ARZ,GTZ,Za} for some important results in 
this other context.
\vskip.3truecm
{\it An outlook of previous results (mostly from a mathematical physics perspective)}.\\
1) In 1971, J. Lebowitz and G. Gallavotti \cite{LG} proved the existence of orientational order for a system of rods on a 
two-dimensional lattice. {\it However}, in their model orientational comes with positional order, which is not the case 
in nematic liquids. In this sense, the system studied by Lebowitz and Gallavotti is 
a model 
for a polarized crystal rather than for a liquid crystal. \\
2) A few months after Lebowitz-Gallavotti's paper, D. Ruelle \cite{Ru71} succeeded in extending the use of the Peierls' method
to a system of infinitely thin rods in the 2D continuum (Widom-Rowlinson model).
Such model exhibits orientational without positional order. In this respect, it represents the first rigorous confirmation 
of the Onsager's excluded volume effect in a system with discrete orientations. 
{\it However}, the assumption of infinitesimal size is unphysical and drastically simplifies the 
analysis: horizontal rods interact only with horizontal rods, and similarly for vertical. It is then natural 
to ask whether the same result remains valid for finite-size molecules. \\
3) The simplest model to look at for studying finite size molecules 
is the $k$-mer model on $\mathbb{Z}^2$: particles are modelled as $1\times k$ rectangles placed on 
the sites of $\mathbb{Z}^2$, either horizontally or vertically, interacting via a hard core repulsion; 
the density of $k$-mers is fixed and smaller than the close packing density (i.e., the rods do not cover the 
plane, but leave a prefixed fraction of the total volume empty). If $k=2$ we have dimers at finite density. In 1972, O. Heilmann and E. Lieb \cite{HL1} proved that the thermodynamic and correlation functions of 
dimers on $\mathbb Z^2$ are analytic
for all finite densities. Therefore, hard-core dimers do not exhibit any nematic-like phase.\\
4) In an attempt of constructing a system of finite size anisotropic molecules exhibiting a nematic-like 
behavior, in 1978, O. Heilmann and E. Lieb \cite{HL2} proposed a variant of the dimer system, by adding 
{\it attractive} forces between aligned dimers. By reflection positivity, they proved the existence of 
orientational order. Presumably, orientational order comes {\it without} translational one \cite{L}, as it should in a
nematic phase.  {\it However}, the  attractive forces
completely change the mechanism driving the ordering transition. We are still left with the question of whether 
the excluded volume effect enough to induce order. What about $k$-mers, with $k\ge 3$? 
To put in in Heilmann and Lieb's words:
{\it It is doubtful whether hard rods on a cubic
lattice without any additional interaction do indeed undergo a phase transition.}\\
5) In 2006, D. Ioffe, Y. Velenik and M. Zahradnik \cite{IVZ} proposed a ``polydisperse" variant of the 
$k$-mer model, where rods of all possible lengths are allowed, with statistical weight depending
on $k$. They mapped the model into the 2D Ising model 
and by the exact solution they proved the existence of an isotropic-nematic transition,
which is the image under the mapping of the usual paramagnetic-ferromagnetic transition in the Ising model.
This result strongly suggests that the same transition should take place in the pure $k$-mer system, 
but unfortunately the method of \cite{IVZ} breaks down in the presence of apparently harmless changes in the relative rods' weights.\\
6) In 2007, A. Ghosh and D. Dhar \cite{GD} numerically identified in a very clear manner 
a nematic phase for pure $k$-mers with $k\ge 7$ at intermediate densities. Their result further confirmed the expectation 
arising from the work of \cite{IVZ}.

\section{The model and the main results}

Motivated by these results, M. Disertori and I recently reconsidered the $k$-mer problem 
on the two-dimensional square lattice \cite{DG}. An informal statement of our main theorem is the following. 
Let $\rho\in(0,1/k)$ be the density of rods.
\vskip.2truecm
{\it For $k$ large enough, if $k^{-2}\ll \rho \ll k^{-1}$, 
the system admits two distinct  infinite volume Gibbs states, characterized by long 
range orientational order (either horizontal or vertical) and no translational order, 
selected  by the boundary conditions.}
\vskip.2truecm
This result is the first rigorous proof of the existence of a nematic phase (i.e., a phase that breaks rotational 
symmetry but does not break translational symmetry) and of the validity of the Onsager's volume effect 
in a system of anisotropic particles of finite fixed size. It solves a conjecture posed more than three decades ago,
as explained in the historical outline above. Before I give a sketch of the proof, let me give a more formal 
statement of the same result, which makes clear how the boundary conditions are chosen, and clarifies certain
key properties of the limiting Gibbs states. 
Even if more formal, the definitions and statements below will still be 
slightly unprecise, in order to avoid cumbersome notations or digressions; for a mathematically precise exposition, I 
refer to \cite{DG}. 

Consider a box $\Lambda\subset \mathbb Z^2$ of side $L$ {\small (eventually, $L\to\infty$)}. We denote by 
$\Omega_\Lambda$ the set of allowed
rod configurations in $\Lambda$ (we shall say that a rod belongs to $\Lambda$ if its geometrical center do so; 
similarly, a rods' configuration belongs to $\Lambda$ if all its rods belong to $\Lambda$). 
Let also $\Omega^+_\Lambda$ (resp. $\Omega^-_\Lambda$) be the set of 
horizontal (resp. vertical) allowed rod configurations in $\Lambda$. We define the grand canonical 
partition function with {\it open boundary conditions} as:
$$Z_0(\Lambda)=\sum_{R\in \Omega_{\Lambda}} z^{|R|}$$
where $z$ is the rods' activity (independent of the orientation) and $|R|$ is the number of rods in the configuration 
$R$. Similarly, the partition function with {\it $+$ boundary conditions} is:
$$Z(\Lambda|+)=\sum_{\substack{R\in \Omega_{\Lambda}:\\
R|_{P^{int}_{\Lambda}}\in\Omega^+_{P^{int}_{\Lambda}}}} z^{|R|} $$
where $P^{int}_\Lambda$ is the internal peel of $\Lambda$, i.e., a $2k$-thick boundary layer of $\Lambda$.
In other words, the partition sum with $+$ boundary conditions is such that all the rods with centers closer than $2k$
to the boundary are forced to be horizontal, while those in the ``bulk" have no constraint on their orientation. An 
analogous definition is valid for $-$ boundary conditions, with the role of the horizontal rods exchanged 
with the vertical. 

In addition to the partition functions, we can define as usual the correlation functions with open, or $+$, or $-$
boundary conditions. For example, the correlation functions with $+$ boundary conditions are defined as follows.
Let $A_X$ be a local observable with finite support $X$, then:
$$ \langle A_X\rangle_\Lambda^{+}=\frac1{Z(\Lambda|+)}
\sum_{\substack{R\in \Omega_{\Lambda}:\\
R|_{P^{int}_{\Lambda}}\in\Omega^+_{P^{int}_{\Lambda}}}}
z^{|R|}\, A_X(R)$$
and $ \langle A_X\rangle^+=\lim\limits_{|\Lambda|\to\infty} \langle A_X\rangle_\Lambda^{+}$. 
The collection of the limiting statistical averages $\langle A_X\rangle_\Lambda^{+}$ as $A_X$ spans a complete set of observables defines the infinite volume Gibbs state with $+$ boundary conditions, 
$\langle \cdot\rangle_\Lambda^{+}$. 
Analogous definitions hold for $\langle\cdot\rangle^-$ or $\langle\cdot\rangle^0$.
In terms of these definitions, a more precise way of formulating our main result is the following.

{\bf Theorem \cite{DG}.} {\it There exist $K_0,\varepsilon_0>0$ such that, if $k\ge K_0$ and $\max\{zk, e^{-zk^2}\}\le 
\varepsilon_0$, then 
the two infinite volume states $\langle\cdot\rangle^\pm$ exist. They are translationally invariant,
pure\footnote{A translationally invariant Gibbs state $\langle\cdot\rangle$ on $\mathbb Z^2$
is said to be pure if it satisfies the cluster property, that is $\lim_{|a|\to\infty}\langle A_XB_{Y+a}\rangle = 
\langle A_X\rangle\langle B_Y\rangle$, where $Y+a$ is the translate of the set $Y\subset \mathbb Z^2$ 
by the vector $a\in\mathbb{Z}^2$.} and are different among each other. In particular:\\
1) If $E^-_{\xi_0}$ is the event that the rods in a window $\Delta_{\xi_0}$ of side $\ell=k/2$ centered at $\xi_0$
are all vertical, then 
$${\rm Prob}^+(E^-_{\xi_0}):=\langle\chi(E^{-}_{\xi_0})\rangle^+\le e^{-c z k^{2}}\;,$$
where $\chi(E^{-}_{\xi_0})$ is the characteristic function of the event and $c$ is a suitable positive constant.\\
2)  If $n_{x}$ is the indicator function that is equal to 
$1$ if a rod has a center in $x\in\mathbb Z^2$ and $0$ otherwise, then
$$ \rho=\langle n_{x}\rangle^+=\langle n_{x}\rangle^-=z(1+O(\varepsilon^{c'}))$$
$$ \rho(x-y)=\langle n_{x}n_{y}\rangle^+=\langle n_{x}n_{y}\rangle^-=
\rho^2\Big(1+O(\varepsilon^{c'|x-y|/k})\Big)$$
where $\varepsilon=\max\{zk,e^{-zk^2}\}$ and $c'$ is a suitable positive constant.}

\section{Ideas of the proof}
While we refer to \cite{DG} for a complete proof, we sketch here the main ideas involved in 
the proof, and its main steps, just to give a flavor of the methods involved.\\
1) The first step consists in coarse graining  $\Lambda $
in square tiles $\Delta$ of side $\ell= k/2$: this means that we partition the original box in 
a collection of tiles, each of which contains in average many ($ \sim z k^2\gg1$) rods; in this respect the 
tiles can be thought of as being mesoscopic.
On the other hand, each tile $\Delta$ is so small (the side is half the rods' length) 
that only rods of the same orientation can have centers 
in $\Delta$, due to the hard rod condition.\\
2) Given a tile $\Delta$, once we prescribe the orientation of the rods with centers in it, the effective interaction between rods 
of the same orientation, say horizontal, is 
weak: in fact, the hard core 
repulsion just prevents two rods to occupy the same row, an event that is very rare, 
since the density of occupied rows is $\sim \frac{z k^2}k\ll1$. It is remarkable that the standard 
{\it cluster expansion} allows us to 
quantify how close to 
Poissonian is the distribution of centers in $\Delta$, once the orientation is prescribed.\\
3) Each tile can thus be of three types: $+1$ (horizontal), $-1$ (vertical) and $0$ (empty). In this way, we can 
associate with every allowed rods' configuration on $\Lambda$ a corresponding
{\it spin configuration} on the coarser lattice of the tiles' centers. By summing over all the rods' configuration
corresponding to a given spin configuration, we are left with a partition sum over spins, which defines an effective spin model. This spin system has the following features: (i)
the interaction between spins is short range and strongly attractive, due to the hard core; (ii)
The vacuum configurations are unlikely, since the probability that $\Delta$ is empty is $\sim e^{-({\rm const.})z k^2}$.
Therefore, the typical spin
configurations consist of big connected clusters of ``uniformly magnetized spins", 
separated by boundary layers (the contours), which contain 
zeros or pairs of neighboring opposite spins.\\ 
4) We are left with studying this effective contour theory. The idea is to use a Peierls argument or cluster expansion methods. However, there are a few issues that make life complicated.\\
(4.a) First of all, the inter-contour interaction 
turns out to be an exponentially decaying $N$-body interaction, with $N$ arbitrary; if the nature of this $N$-body 
interaction were generic then we would really be in trouble: nobody knows how to work out a convergent cluster 
expansion for generic $N$-body interactions, even if exponentially decaying. Luckily enough, 
the $N$-body interaction we need to deal with is quasi-1D, in the sense that 
only contours whose horizontal or vertical projections overlap interact among each other. This makes the 
problem treatable. The right strategy is to follow Brydges' suggestion \cite{Bry}: {\it if at first you do not
succeed, then expand and expand again!} In other words, we perform a second Mayer expansion of the 
multi-contour interaction and we collect together the connected components (polymers); the resulting polymers have purely hard core 
interactions, and the polymers' activities satisfy similar bounds as those of the original contours. \\
(4.b)
The activity of the contour is defined in terms of a ratio of partition functions that at the beginning of the story 
we do not know how to compute (yet). Moreover, the contour theory is not exactly 
symmetric under spin flip, due to the finite size of the rods (i.e., the activity of a given contour is not the same in the presence of $+$ or $-$ boundary conditions, unless the shape of the contour itself is rotation invariant).
To solve these two issues, we use the methods of Pirogov-Sinai theory \cite{PS,Z}.
By induction, we show that the polymers satisfy  a Peierls' 
condition, i.e., the probability that a given contour or polymer occurs is exponentially small in the 
size of its geometric support.
For details, see \cite{DG}.

\section{Conclusions and open problems}

I reported the first proof of the existence of a nematic-like phase in the pure $k$-mer system in two dimensions,
for $k$ sufficiently large and intermediate densities. This proof solves a long standing conjecture about the 
validity of the Onsager's excluded volume effect in systems of finite size anisotropic molecules. Of course,
many questions remain open. Let us mention a few of them.\\
1) {\it Three dimensions.} It would be nice to extend the results of \cite{DG} to three dimensions. 
Presumably, an analogous result should remain valid for $1\times1\times k$ rods on $\mathbb Z^3$. A more
interesting question is what happens for anisotropic molecules of the form $1\times k^\alpha\times k$ ($0<\alpha<1$)
with six allowed orientations and centers in $\mathbb R^3$: in this case a priori there is the possibility that 
other kinds of liquid crystalline order (smectic, chiral, etc \cite{dG}) emerge.\\
2) {\it High density.}
As $\rho$ is increased towards close-packing, a second transition to a ``disordered" 
state (i.e., a state with no orientational order) is expected \cite{GD}. 
The understanding of this dense phase is completely open.
In particular, at close packing, the best known variational bound on the entropy is: 
$\frac{S_{cp}}{|\Lambda|}\ge C_{s}\frac{\log k}{k^2}$ for an explicitly known constant $C_s$,  
which is obtained by making a ``striped" ansatz \cite{GD}. 
A very interesting open problem is to prove that $\frac{S_{cp}}{|\Lambda|}\sim\frac{\log k}{k^2}$, asymptotically 
as $k\to\infty$.\\
3) 
{\it Rotational invariance.}
``Of course", the most important and difficult open problem is to prove the existence of orientational
order in a model with genuine rotational invariance. There are some examples of two or three-dimensional gases
with a continuous internal degree of freedom, interacting via a tensorial {\it attractive} 
interaction, where the existence of nematic order (or quasi-order in 2D) 
can be proved by a combination of cluster expansion 
and reflection positivity \cite{AZ,ARZ,GTZ,Za}. However, the understanding of the Onsager's excluded 
volume effect is completely open in this case. I think that this last problem is substantially more difficult than the previous two. We have a serious lack of understanding of continuous symmetry breaking
phenomena, as witnessed by the state of art in the problem of existence of spontaneous magnetization 
in 3D classical and quantum ferromagnets: there are only very few special cases where 
spontaneous magnetization can be rigorously proved \cite{FSS,DLS}, the proof being based on reflection 
positivity, which is very fragile under apparently harmless changes in the Hamiltonian. It is very likely 
that future progress on the theory of the Heisenberg ferromagnet 
will also help us in understanding the validity of the Onsager's excluded volume effect 
in continuous three dimensional systems. It may also be of help to try to attack the issue 
of continuous rotational symmetry by first looking 
at mean-field like models, possibly in the spirit of \cite{LP}. 
\vskip.5truecm
{\bf Acknowledgements.} The research leading to these results has received
funding from the European Research Council under the European Union's Seventh
Framework Programme ERC Starting Grant CoMBoS (grant agreement n$^o$ 239694).
I would like to thank the committee of the IUPAP Young Scientist prize for the great honor they made me,
and for which I am deeply grateful. I would also like to warmly thank my teachers and collaborators, from which I 
learnt how to make research in mathematical physics, and thanks to which I got here: Giovanni Gallavotti,
Giuseppe Benfatto, Vieri Mastropietro, Joel Lebowitz and Elliott Lieb. Thanks a lot.

\bibliographystyle{ws-procs975x65}

\end{document}